\documentclass[%
 reprint,
 amsmath,amssymb,
 aps
]{revtex4-1}

\usepackage{graphicx}
\usepackage{dcolumn}
\usepackage{bm}
\usepackage[dvipdfm]{hyperref}
\usepackage{color}
\usepackage{setspace}
\begin{document}
\newcommand{\average}[1]{\ensuremath{\langle#1\rangle} }
\renewcommand{\figurename}{Fig}
\renewcommand{\tablename}{Table}
\preprint{APS/123-QED}

\title{Possible effects of collective neutrino oscillations in the three flavor multi-angle simulations on supernova $\nu p$ process}

 

\author{H. Sasaki$^{1,2}$, T. Kajino$^{1,2,3}$, T. Takiwaki$^{2}$, T. Hayakawa$^{2,4}$, A. B. Balantekin$^{2,5}$ and Y. Pehlivan$^{2,6}$}
\affiliation{%
\scriptsize{
$^{1}$Department of Astronomy Graduate School of Science The University of Tokyo, 7-3-1 Hongo, Bunkyo-ku, Tokyo 113-033, Japan\\
}
\scriptsize{
$^{2}$National Astronomical Observatory of Japan, 2-21-1 Osawa, Mitaka, Tokyo 181-8588, Japan\\
}
\scriptsize{
$^{3}$International Research Center for Big-Bang Cosmology and Element Genesis, and School of Physics and Nuclear Energy Engineering, Beihang University, Beijing 100191, P.R. China \\
}
\scriptsize{
$^{4}$National Institutes of Quantum and Radiological Science and Technology, Shitakata 2-4, Tokai, Ibaraki, 319-1106, Japan\\
}
\scriptsize{
$^{5}$Department of Physics, University of Wisconsin, Madison, WI 53706, USA\\
}
\scriptsize{
$^{6}$Department of Physics, Mimar Sinan Fine Arts University, Istanbul 34380, Turkey
}
}%




\setstretch{1.5}
\begin{abstract}
 \ We study the effects of collective neutrino oscillations on $\nu p$ process nucleosynthesis in proton-rich neutrino-driven winds by including both the multi-angle $3\times3$ flavor mixing and the nucleosynthesis network calculation. 
The number flux of energetic electron antineutrinos is raised by collective neutrino oscillations in a $1$D supernova model for $40 M_{\odot}$ progenitor. When the gas temperature decreases down to $\sim2-3\times10^{9}$ K, the increased flux of electron antineutrinos promotes $\nu p$ process more actively, resulting in the enhancement of $p$-nuclei. In the early phase of neutrino-driven wind, blowing at $0.6$ s after core bounce, oscillation effects are prominent in inverted mass hierarchy and $p$-nuclei are synthesized up to $^{106}\mathrm{Cd}$ and $^{108}\mathrm{Cd}$. On the other hand, in the later wind trajectory at $1.1$ s after core bounce, abundances of $p$-nuclei are increased remarkably by $\sim10-10^{4}$ times in normal mass hierarchy and even reaching heavier $p$-nuclei such as $^{124}\mathrm{Xe}$, $^{126}\mathrm{Xe}$ and $^{130}\mathrm{Ba}$. The averaged overproduction factor of $p$-nuclei is dominated by the later wind trajectories. Our results demonstrate that collective neutrino oscillations can strongly influence $\nu p$ process, which indicates that they should be included in the network calculations in order to obtain precise abundances of $p$-nuclei. The conclusions of this paper depend on the difference of initial neutrino parameters between electron and non-electron antineutrino flavors which is large in our case. Further systematic studies on input neutrino physics and wind trajectories are necessary to draw a robust conclusion. However, this finding would help understand the origin of solar-system isotopic abundances of $p$-nuclei such as $^{92,94}\mathrm{Mo}$ and $^{96,98}\mathrm{Ru}$.\\





\end{abstract}

\pacs{Valid PACS appear here}
\maketitle


 \section{Introduction}
\begin{figure}
 \begin{center}
 \rotatebox{0}{
\includegraphics[angle=0,bb=30 0 1930 850,clip,scale=0.45]{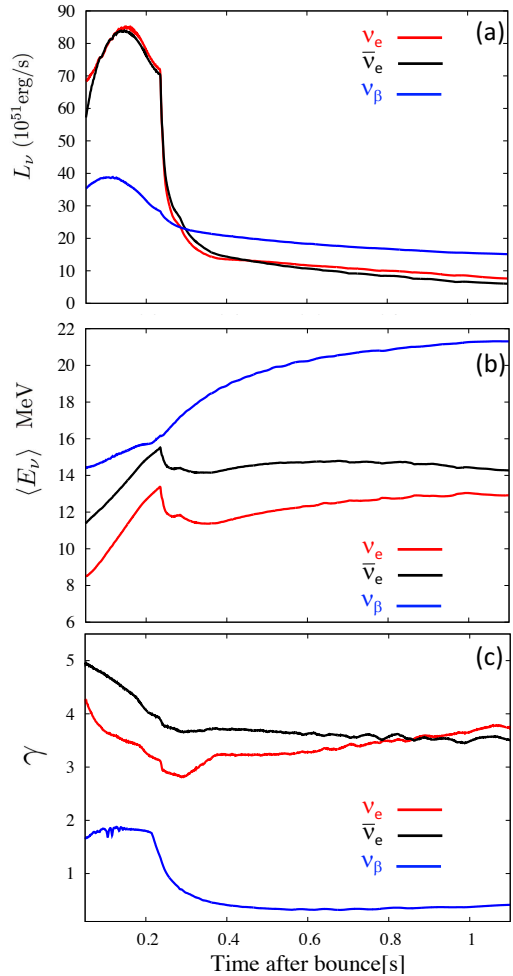}
}
 \caption{The time evolution of neutrino luminosities $L_{\nu}$ (a), mean energies $\average{E_{\nu}}$ (b) and shape parameters $\gamma$ (c) in the $1$D explosion model where $\nu_{\beta}=\nu_{\mu}, \nu_{\tau}, \bar{\nu}_{\mu}, \bar{\nu}_{\tau}$.}\label{FIG:neutrino_parameter_evolution}
 \end{center}
 \end{figure}

 Several cosmological and astrophysical sites such as the Early Universe, the core-collapse supernovae, and neutron star mergers are intense neutrino sources.  In core-collapse supernovae, during $\sim1-10$ seconds after core bounce, $\sim 10^{58}$ neutrinos and antineutrinos ($\nu_{e},\nu_{\mu},\nu_{\tau},\bar{\nu}_{e},\bar{\nu}_{\mu},\bar{\nu}_{\tau}$) are emitted from the proto-neutron star and carry away the gravitational binding energy out of the inner core \cite{Colgate 1966}. At such high neutrino number densities, coherent superposition of neutrino-neutrino scattering amplitudes triggers a self refraction effect which induces dramatic flavor transformation modes as emergent many-body phenomena \cite{Fuller 1987,Pantaleone 1992_1,Pantaleone 1992_2,Raffelt1993,Balantekin 2007,Pehlivan 2011,Pehlivan 2014, Pehlivan 2016}. These are called ``collective neutrino oscillations'' because both analytical and numerical studies indicate that the strong correlations develop between flavor evolution of neutrinos with different momenta  \cite{Duan 2006,Hannestad 2006,Fogli 2007,Esteban-Pretel 2007,Dasgupta 2009,Dasgupta 2010,Duan 2010,Duan 2011,Mirizzi 2011,Chakraborty 2014,Tian 2017}. Collective neutrino oscillations transform the spectra of all neutrino species, but particularly important for our purposes is the modification of $\nu_{e}$ and $\bar{\nu}_{e}$ energy distributions because their absorptions on free nucleons through $\nu_{e}+n\to e^{-}+p$ and $\bar{\nu}_{e}+p\to e^{+}+n$ reactions significantly affect the nucleosynthesis.

\ It was proposed that explosive nucleosynthesis takes place in neutrino-driven winds. Previous numerical studies \cite{Fischer 2010,Hudepohl 2010} suggest that neutrino-driven winds become proton-rich outflows ($Y_{e}>0.5$) rather than neutron-rich outflows ($Y_{e}<0.5$), where $Y_{e}$ is the electron fraction inside the outflow. 
The $\nu p$ process \cite{Frohlich 2006,Pruet 2006,Wanajo 2006} is proposed as a primary nucleosynthesis induced by free protons and neutrons supplied by the $p$($\bar{\nu}_{e}$,$e^{+}$)$n$ interactions in proton-rich outflows.  
These free neutrons allow the creation of heavier elements beyond the waiting point nucleus $^{64}\mathrm{Ge}$ via $^{64}\mathrm{Ge}(n,p)^{64}\mathrm{Ga}$ instead of $\beta^{+}$ decay.
The $\nu p$ process can synthesize $p$-nuclei which are located in the proton-rich side of stability line and bypassed by the major two neutron capture reactions of $r$- and $s$-processes.
\\

\ In proton-rich outflows, increased $\bar{\nu}_{e}$ flux induced by collective neutrino oscillations may enhance the $\nu p$ process. Conversely, the abundances of the affected nuclides may be used as a probe to investigate non-linear effects of collective neutrino oscillations on the neutrino spectra in addition to direct measurements of neutrino fluxes.\\

\ The effects of collective neutrino oscillations on nucleosynthesis have been considered in the previous studies \cite{Balantekin 2005,Duan_nucleo 2011,Martinez 2011,Tamborra 2012,Pllumbi 2015,Wu 2015}. In neutron-rich outflows, it was reported that the use of single-angle approximation \cite{Duan 2006} leads to inaccurate prediction for the yields \cite{Duan_nucleo 2011}. This is because the single-angle approximation ignores the angular dependence of emitted neutrinos and causes an early onset of collective flavor transformations \cite{Duan 2011}. In the multi-angle calculation \cite{Duan 2006,Fogli 2007,Duan 2011,Mirizzi 2011,Chakraborty 2014,Duan_nucleo 2011,Wu 2015}, however, the angular dependence of flavor evolution is taken into account and oscillation phenomena can be predicted more realistically.\\

\ In proton-rich outflows, it was shown that when spectral swaps caused by collective neutrino oscillations are systematically included, the abundances of $p$-nuclei are enhanced \cite{Martinez 2011}. However, the simple spectral split scenario adopted in Ref. \cite{Martinez 2011} does not always occur in collective neutrino oscillations. A realistic calculation which couples collective neutrino oscillations with nucleosynthesis network calculations has not yet been carried out in proton-rich outflows. Such treatment is required because of the difficulty to predict the onset of collective neutrino oscillations which plays significant roles in the nucleosynthesis.\\



\ In this work, we study the impact of collective neutrino oscillations on the $\nu p$ process by combining three flavor and multi-angle simulations for the first time with nucleosynthesis network calculations based on a spherically symmetric $1$D explosion model of a core-collapse supernova. \\

\ This paper is organized as follow:
In section \ref{sec:Setup for numerical simulations}, we introduce the setup for our simulations. In section \ref{sec:Results}, we present the calculated simulation results of oscillation phenomena and their influence on $\nu p$ process nucleosynthesis in both early and later neutrino-driven winds. Discussions about the obtained results and summary in this work are presented in section \ref{sec:Summary and Discussion}.\\
\section{Setup for numerical simulations}
\label{sec:Setup for numerical simulations}
 
 \ We employ $1$D wind models based on the time-dependent neutrino radiation hydrodynamic simulation. The numerical setup is similar to that of Ref. \cite{Sotani 2016} except for the inclusion of phenomenological general relativistic effects on the gravitational potential \cite{Marek 2006}. As the initial profile of the simulation, $40 M_{\odot}$ progenitor model in Ref. \cite{Woosley 2002} is used.
 To obtain a shock revival in $1$D, we reduce the mass accretion rate as in Ref. \cite{Hudepohl 2013}. Fig.\ref{FIG:neutrino_parameter_evolution} represents the time evolution of neutrino luminosities $L_{\nu}$, mean energies $\average{E_{\nu}}$ and shape parameters $\gamma$ (see Eq.(\ref{eq:gamma}) and note that $\alpha$ is often used in other references $e.g.$ Ref. \cite{Tamborra 2014}) in this explosion model. The sharp deacrease of the luminositites at $t=250$ ms after bounce originates from the sudden decrease of the mass accretion rate. Basically, it corresponds to the arrival of the Si layer to the shock. In this work, the accretion rate is reduced by hand and the shock revives at that time. In the late phase, the mean energy of $\nu_{\beta}$ is higher than that reported in recent sophisticated simulations $e.g.$ Ref. \cite{Mirizzi 2016} since inelastic effect of neutrino nucleon down scattering is not taken into account in our simulation (see Fig.$14$ in Ref. \cite{Muller 2012}).\\
 
 \ We choose two representative wind trajectories at $t=0.6$ s and $1.1$ s after core bounce as the fiducial models in the cooling phase.
Neutrino oscillations and nucleosynthesis are calculated as post processes using these wind models from $r=40-3300$ km where $r$ is the distance from the center. The electron fraction inside the outflow is given by
\begin{equation}
\label{eq:ye}
Y_{e}=\sum_{i=\mathrm{all\ species}}\frac{Z_{i}}{A_{i}}X_{i},
\end{equation}
where $Z_{i}$, $A_{i}$ and $X_{i}$ denote atomic number, mass number and mass fraction of nuclear species $i$, respectively. In the cooling phase, the feedback effect of neutrino oscillations on $Y_{e}$ is negligible at $r>100$ km, where collective neutrino oscillations occur, for the following two reasons. The first reason is that in our wind model, the outflow velocity $v(r)$ is so fast that the feedback effect of collective neutrino oscillations does not change the value of $Y_{e}$ remarkably. The second reason is that few free nucleons are produced by $n$($\nu_{e}$,$e^{-}$)$p$ and $p$($\bar{\nu}_{e}$,$e^{+}$)$n$ even though oscillation effects are taken into account. Such a small amount of free nucleons fails to alter the value of $Y_{e}$ sufficiently. As the gas temperature $T$ decreases, large numbers of free nucleons are consumed in the $\alpha$-particle creation, so that target nucleons for the neutrino-induced reactions are exhausted. In our wind models at $t=0.6\ (1.1)\ \mathrm{s}$, the electron fraction inside the outflow actually takes nearly the same constant value $Y_{e}\sim0.55\ (0.59)$ in $r>40$ km independent of neutrino oscillation effects.

\ Neutrino reaction rates for 
$\nu_{e}+n\to e^{-}+p$ and $\bar{\nu}_{e}+p\to e^{+}+n$
are estimated by using the analytical cross sections \cite{Balantekin 2005}
\begin{equation}
\sigma_{\nu_{e}}=9.6\times10^{-44}(E/\mathrm{MeV}+1.293)^{2},
\end{equation}
and
\begin{equation}
\label{eq:cross section nube}
\sigma_{\bar{\nu}_{e}}=9.6\times10^{-44}(E/\mathrm{MeV}-1.293)^{2}\ \mathrm{cm}^{2},
\end{equation}
respectively. We include not only neutrino absorptions on free nucleons but also the electron and positron capture reactions \cite{Tamborra 2012}, and neutrino absorptions on $\alpha$-particles as discussed in Ref. \cite{Terasawa 2004,Yoshida 2008}. Cross sections of the $\alpha$-induced reactions, derived by the WBP Hamiltonian \cite{Warburton 1992} are no longer negligible because $\alpha$-particles become dominant species in neutrino-driven winds after the wind temperature decreases down to $T\sim 6\times10^{9}$ K.
The data of other nuclear reaction rates on more than $8000$ nuclides are adopted from JINA Reaclib database \cite{Cyburt 2010}. Nucleosynthesis in neutrino-driven winds is calculated by running $\textit{libnucnet}$ reaction network engine \cite{Meyer 2007}. The effects of neutrino oscillations are included in the network calculation successively.\\


 \ We adopt the following neutrino oscillation parameters in our simulations: $\theta_{23}=45^\circ$, $\theta_{13}=8.5^\circ$, $\theta_{12}=34^\circ$, $\Delta m^{2}_{21}=7.5\times10^{-5} \mathrm{eV}^{2}$, $|\Delta m^{2}_{32}|=2.4\times10^{-3} \mathrm{eV}^{2}$ and $\delta_{CP}=0$ where $\Delta m^{2}_{ij}=m^{2}_{i}-m^{2}_{j}$. The positive (negative) $\Delta m^{2}_{32}$ defines normal (inverted) mass hierarchy, respectively. We set the same radius of neutrino sphere $R_{\nu}=18$ km irrespective of neutrino species in both wind models at $t=0.6$ s and $1.1$ s. This assumption is applicable to our calculation because the onset radius of collective neutrino oscillations  \cite{Duan 2011} is not sensitive to a small difference by few km in $R_{\nu}$. On the surface of the neutrino sphere ($r=R_{\nu}$), we impose the normalized neutrino spectra $f_{\nu_{\alpha}}$ (for $\alpha=e, \mu, \tau$) \cite{Keil 2003}
 \begin{equation}
f_{\nu_{\alpha}}(E)=\frac{E^{\gamma}}{\Gamma(\gamma+1)}\left(
 \frac{\gamma+1}{\average{E_{\nu_{\alpha}}}}
 \right)^{\gamma+1}\exp\left[
 -\frac{(\gamma+1)E}{\average{E_{\nu_{\alpha}}}}
 \right],
 \end{equation}
 with, 
 
 \begin{equation}
 \label{eq:gamma}
 \gamma=\frac{\average{E_{\nu_{\alpha}}^{2}}-2\average{E_{\nu_{\alpha}}}^{2}}{\average{E_{\nu_{\alpha}}}^{2}-\average{E_{\nu_{\alpha}}^{2}}},
 \end{equation}

\noindent where $\gamma$ is a shape parameter and $\Gamma(x)$ is the gamma function. The normalized antineutrino spectra $f_{\bar{\nu}_{\alpha}}$ (for $\alpha=e, \mu, \tau$) are also introduced in the same way. Table \ref{tab:multi-angle parameters} shows the initial neutrino parameters in our models obtained by the 1D explosion simulation. From the radius of neutrino sphere ($r=R_{\nu}$) to the beginning of the oscillation calculation ($r=40$ km), we neglect any flavor transitions because of the presence of dominant matter effects and the multi-angle decoherence \cite{Duan 2011,Mirizzi 2011,Duan_nucleo 2011}.\\
\begin{table*}[htb]
\begin{center}
\small
\begin{tabular}{|cccccccccc|} \hline

$t$&$L_{\nu_{e}}$&$L_{\bar{\nu}_{e}}$&$L_{\nu_{\beta}}$&$\average{E_{\nu_{e}}}$&$\average{E_{\bar{\nu}_{e}}}$&$\average{E_{\nu_{\beta}}}$&$\gamma_{\nu_{e}}$&$\gamma_{\bar{\nu}_{e}}$&$\gamma_{\nu_{\beta}}$\\ 
(s)&$(10^{51}$erg/s)&$(10^{51}$erg/s)&$(10^{51}$erg/s)&(MeV)&(MeV)&(MeV)&&&\\ \hline
 $0.6$&$\ \ 11.7$&$\ \ 10.7$&$\ \ 18.3$&$\ \ 12.3$&$\ \ 14.7$&$\ \ 20.2$&$\ \  3.16$&$\ \ 3.66$&$\ \ 0.32$\\ 
 $1.1$&$\ \ 7.6$&$\ \ 6.0$&$\ \ 15.1$&$\ \ 12.9$&$\ \ 14.3$&$\ \ 21.3$&$\ \  3.72$&$\ \ 3.53$&$\ \ 0.42$\\ \hline
\end{tabular}
\end{center}
\small
\caption{The parameter set of neutrinos on the surface of the neutrino sphere where $\nu_{\beta}=\nu_{\mu}, \nu_{\tau}, \bar{\nu}_{\mu}, \bar{\nu}_{\tau}$.}\label{tab:multi-angle parameters}
\end{table*}

\ We perform the three flavor multi-angle calculations by employing the neutrino ``bulb model'' \cite{Duan 2006}. In this treatment, flavor contents of emitted neutrinos can be represented by a $3\times3$ density matrix $\rho(r,E,\theta_{p})$ where $E$ is neutrino energy, and $\theta_{p}$ is the angle of the neutrino propagation direction with respect to the radial direction. The corresponding density matrix for antineutrinos is denoted by $\bar{\rho}(r,E,\theta_{p})$. We normalize the traces of $\rho$ and $\bar{\rho}$ as $\mathrm{Tr}\rho=\mathrm{Tr}\bar{\rho}=1$, which allows to impose a probabilistic interpretation on the diagonal components, e.g., the $\rho_{\alpha\alpha}(r,E,\theta_{p})$ is the probability of finding a neutrino in $\alpha$-flavor with energy $E$, propagating in direction of $\theta_{p}$ at a distance $r$ from the center. The three flavor, multi-angle calculation is carried out by solving the equations of motions of neutrino and antineutrino density matrices \cite{Raffelt1993}
\begin{eqnarray}
\label{eq:time evolution neutrino r}
\cos &\theta_{p}&\frac{\partial }{\partial r}\rho(r,E,\theta_{p})  \\ \nonumber &=& -i
\left[\rho(r,E,\theta_{p}),\Omega(E)+V(r,E,\theta_{p})
\right],
\end{eqnarray}

\begin{eqnarray}
\label{eq:time evolution antineutrino r}
\cos &\theta_{p}& \frac{\partial }{\partial r}\bar{\rho}(r,E,\theta_{p}) \nonumber \\ &=& -i
\left[\bar{\rho}(r,E,\theta_{p}),-\Omega(E)+V(r,E,\theta_{p})
\right].
\end{eqnarray}
Here $\Omega(E)$ is the vacuum oscillation Hamiltonian 
\begin{eqnarray}
\Omega(E)=\frac{\Delta m^{2}_{21}}{6E}U\left(
\begin{array}{c c c}
-2&0&0\\
0&1&0\\
0&0&1\\
\end{array}\right)U^{\dagger}\\ \nonumber 
+\frac{\Delta m^{2}_{32}}{6E}U\left(
\begin{array}{c c c}
-1&0&0\\
0&-1&0\\
0&0&2\\
\end{array}\right)U^{\dagger},
\end{eqnarray}
where $U$ is the Pontecorvo-Maki-Nakagawa-Sakata (PMNS) matrix \cite{Maki 1962} which includes the mixing angles $\theta_{ij}$. The potential consists of two terms: $V(r,E,\theta_{p})=V_{\mathrm{matter}}(r)+V_{\mathrm{self}}(r,\theta_{p})$. Here $V_{\mathrm{matter}}(r)$ represents the effect of the net electron background \cite{Wolfenstein 1978,Mikheyev 1985}
\begin{equation}
V_{\mathrm{matter}}(r)=\sqrt{2}G_{F}n_{e}(r)\left(
\begin{array}{c c c}
1&0&0\\
0&0&0\\
0&0&0\\
\end{array}\right),
\end{equation}
where $n_{e}(r)$ is the electron density in the radius $r$, and $V_{\mathrm{self}}(r,\theta_{p})$ is the potential of neutrino self interactions whose strength is determined by neutrino luminosities \cite{Raffelt1993,Duan 2006}. It is given by
\begin{eqnarray}
\label{eq:multi potential}
V_{\mathrm{self}}(r,E,\theta_{p})=\frac{\sqrt{2}G_{F}}{2\pi R_{\nu}^{2}}\int \mathrm{d}E\ \mathrm{d}(\cos\theta_{q})
(1-\cos\theta_{p}\cos\theta_{q}) \nonumber \\ \sum_{\alpha=e,\mu,\tau} \left\{
\frac{L_{\nu_{\alpha}}}{\average{E_{\nu_{\alpha}}}}f_{\nu_{\alpha}}(E)\rho(r,E,\theta_{q})-\frac{L_{\bar{\nu}_{\alpha}}}{\average{E_{\bar{\nu}_{\alpha}}}}f_{\bar{\nu}_{\alpha}}(E)\bar{\rho}(r,E,\theta_{q})\right\}. \nonumber  \\
\end{eqnarray}
In our calculations, we adopt the mean-field approach and ignore any sterile neutrino mixings. \\
\begin{figure*}
 \begin{center}
 \rotatebox{-90}{
 \includegraphics[angle=0,bb=-5 0 530 840,clip,scale=0.5]{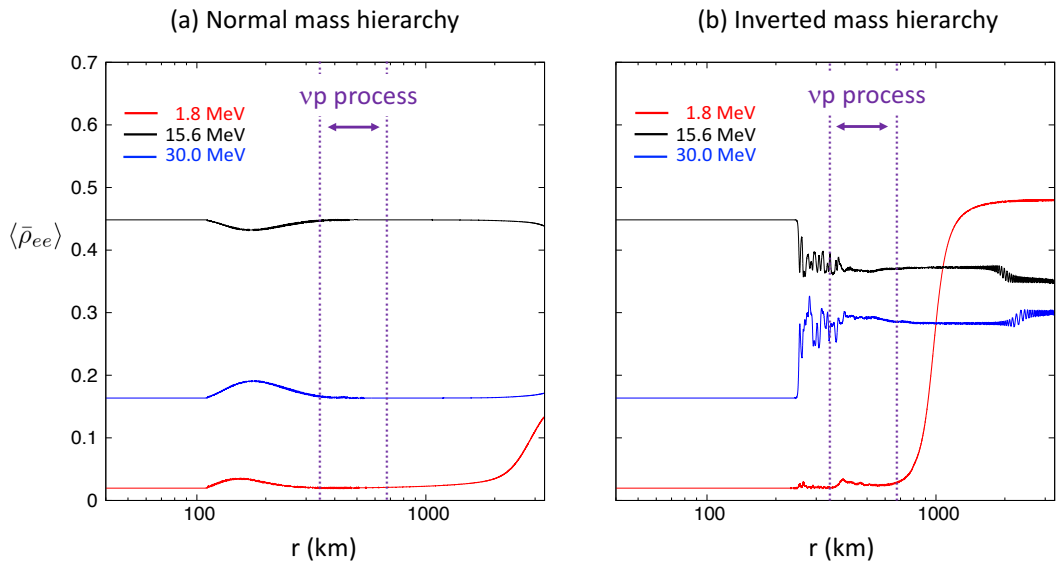}
 }
 \caption{This figure shows the evolution of angle averaged \average{\bar{\rho}_{ee}(r,E)} in normal (a) and inverted (b) mass hierarchies in the early flow ($t=0.6$ s). \average{\bar{\rho}_{ee}(r,E)} represents the ratio of $\bar{\nu}_{e}$ in an antineutrino whose energy is $E$ at $r$. The $\nu p$ process takes place during $T\sim2-3\times10^{9}$ K which corresponds to $r\sim350-680$ km.
 }\label{FIG:make_rho_be_evolution_3_m_0.6}
 \end{center}
 \end{figure*}


\section{Results}
\label{sec:Results}


\ Here we first present numerical results of collective neutrino oscillations and their influence on neutrino-induced reaction rates in the early wind ($t=0.6$ s) in section \ref{subsec:The early trajectory $(t=0.6$ s$)$}. Results in the later outflow ($t=1.1$ s) are discussed in section \ref{subsec:The later trajectory $(t=1.1$ s$)$}. Finally, effects of oscillations on abundances of $p$-nuclei are discussed in section \ref{subsec:The abundances of $p$-nuclei produced in neutrino-driven winds}.





\begin{figure}
 \begin{center}
 \rotatebox{0}{
\includegraphics[angle=0,bb=30 0 1930 800,clip,scale=0.45]{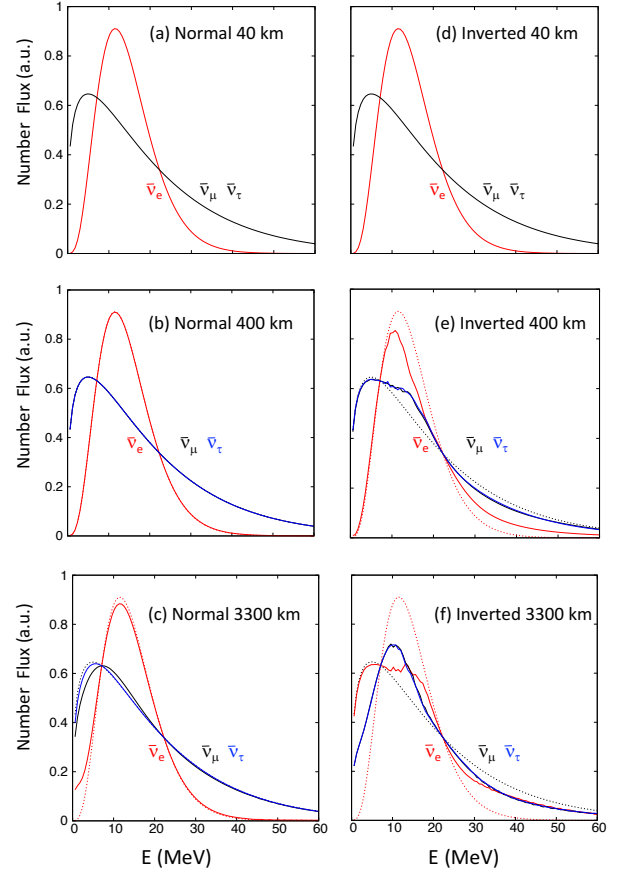}
}
 \caption{The evolution of energy spectra of antineutrinos from $r=40$ km to $3300$ km in both normal and inverted mass hierarchies using the early wind trajectory ($t=0.6$ s). Thin dashed curves display initial antineutrino spectra. There are two spectral crossing points in antineutrino spectra whose energies are $E_{c1}^{(e)}=7.1$ MeV and $E_{c2}^{(e)}=22.3$ MeV.}\label{FIG:make_spec_0.6}
 \end{center}
 \end{figure}

\subsection{Early neutrino-driven wind $(t=0.6$ s$)$} 
\label{subsec:The early trajectory $(t=0.6$ s$)$}


\ Collective neutrino oscillations are caused by the non-linear self-interacting potential $V_{\mathrm{self}}(r,\theta_{p})$. These oscillations affect energy spectra of all species of neutrinos. In inverted mass hierarchy, the $\nu p$ process is enhanced by the increased number of energetic electron antineutrinos.\\


\ Fig.\ref{FIG:make_rho_be_evolution_3_m_0.6} shows the evolution of the angle averaged ratio of electron antineutrinos for the early wind model of $t=0.6$ s given by
\begin{equation}
\average{\bar{\rho}_{ee}(r,E)}=\frac{2}{\pi}\int_{0}^{\frac{\pi}{2}}\mathrm{d}\theta_{R}\ \bar{\rho}_{ee}(r,E,\theta_{p}),
\end{equation}
for the three typical energies $1.8$, $15.6$ and $30$ MeV. Here $\theta_{R}$ is the emission angle on the surface of the neutrino sphere which is in one-to-one relation to $\theta_{p}$ \cite{Duan 2006}. Our results for normal and inverted mass hierarchies are shown in
Fig.\ref{FIG:make_rho_be_evolution_3_m_0.6}(a) 
and \ref{FIG:make_rho_be_evolution_3_m_0.6}(b), respectively. The evolution of the full antineutrino energy spectra is shown in Fig.\ref{FIG:make_spec_0.6} where the left (right) column corresponds to normal (inverted) mass hierarchy.\\


Following Ref. \cite{Balantekin:1999dx}, we introduce the combinations
\begin{equation}
\nu_x = \cos  \theta_{23} \> \nu_{\mu} - \sin  \theta_{23} \> \nu_{\tau},
\end{equation}
\begin{equation}
\nu_y = \sin  \theta_{23} \> \nu_{\mu} + \cos  \theta_{23} \> \nu_{\tau}. 
\end{equation}
In normal mass hierarchy (Fig.\ref{FIG:make_rho_be_evolution_3_m_0.6}(a)), the synchronization
due to the neutrino self interactions \cite{Hannestad 2006,Fogli 2007} and
high electron density prevent any flavor transitions until $r\sim 110$ km. Henceforth, the decreasing neutrino self interaction potential becomes
comparable to the vacuum oscillation term and
all neutrino species begin to change flavor collectively,
irrespective of their momenta and direction of motion. In such collective phenomena, ${\bar{\nu}}_e- \bar{{\nu}}_y$ conversions \cite{Dasgupta 2010, Mirizzi 2011}  occur and $\bar{\nu}_{x}$ is decoupled from other flavor of antineutrinos because $\bar{\nu}_{\mu}$ and $\bar{\nu}_{\tau}$ acquire about the same effective mass inside the dense material \cite{Balantekin:1999dx}. 


Around $r=400$ km where the $\nu p$ process takes place, the contribution of $V_{\mathrm{self}}(r,\theta_{p})$ in the total neutrino Hamiltonian is negligible, so that collective neutrino oscillations have terminated. As shown in Fig.\ref{FIG:make_spec_0.6}(b), any spectral swaps can not be observed because affected antineutrinos have finally come back to their original flavors in the end of collective neutrino oscillations. This implies that the effects of neutrino oscillations on the neutrino-induced reactions are negligible in normal mass hierarchy. After that, low energy antineutrinos start changing flavor gradually as shown, for example, by the $1.8$ MeV antineutrinos in Fig.\ref{FIG:make_rho_be_evolution_3_m_0.6}(a). Decreasing electron density allows low energy antineutrinos to couple with the solar vacuum frequency $\omega_{\mathrm{solar}}=\Delta m^{2}_{21}/2E$ resulting in the adiabatic neutrino flavor transitions to the vacuum mass eigenstates. This matter effect causes the difference between $\bar{\nu}_{\mu}$ flux and that of $\bar{\nu}_{\tau}$ as shown in Fig.\ref{FIG:make_spec_0.6}(c) because vacuum mass eigenstates are combinations of flavor eigenstates via the PMNS matrix $U$.\\ 


\ In inverted mass hierarchy (Fig.\ref{FIG:make_rho_be_evolution_3_m_0.6}(b)), collective neutrino oscillations start around $r=250$ km, which results in the transformation of energetic electron antineutrinos around $r=400$ km as shown, for example by the $15.6$ and $30$ MeV antineutrinos in Fig.\ref{FIG:make_rho_be_evolution_3_m_0.6}(b). $\bar{\nu}_{\mu}$ and $\bar{\nu}_{\tau}$ are almost degenerate during the collective neutrino oscillations, resulting in the same energy spectra as shown in Fig.\ref{FIG:make_spec_0.6}(e). The spectral splits caused by collective neutrino oscillations develop around the spectral crossing points in antineutrino spectra \cite{Dasgupta 2009,Dasgupta 2010}. Flavor transitions are observed in $E>E_{c1}^{(e)}=7.1$ MeV. Here $E_{c1}^{(e)}$ represents the value of the first spectral crossing point in our antineutrino spectra. The flavor transitions of low energy antineutrinos ($E<E_{c1}^{(e)}$) are highly suppressed because of the multi-angle decoherence. The increased number of electron antineutrinos whose energies are larger than the value of the second spectral crossing point ($E_{c2}^{(e)}=22.3$ MeV) cause the enhancement of the $\nu p$ process nucleosynthesis. 
Complete spectrum swaps as obtained in calculations with the single-angle approximation do not emerge from our multi-angle calculations. This smeared oscillation phenomenon is consistent with the previous numerical studies \cite{Fogli 2007,Mirizzi 2011}.  After the collective neutrino oscillations cease, antineutrinos undergo Mikheyev-Smirnov-Wolfenstein (MSW) resonances \cite{Wolfenstein 1978,Mikheyev 1985} between $\bar{\nu}_{e}$ and $\bar{\nu}_{y}$ caused by the coupling between the atmospheric vacuum frequency $\omega_{\mathrm{atm}}=\Delta m^{2}_{32}/2E$ and the matter potential $V_{\mathrm{matter}}(r)$.\\



\  In both hierarchies, the onset of collective neutrino oscillations are delayed compared with that in single-angle approximation. Such delayed collective neutrino oscillations are caused by the angular dispersion of $V_{\mathrm{self}}(r,\theta_{p})$ as discussed in \cite{Duan 2011}. These multi-angle effects make critical deviations in
nucleosynthesis yields inside the neutrino-driven winds
in comparison to the 
single-angle calculations \cite{Duan_nucleo 2011}. The use of single angle approximation would start collective neutrino oscillations earlier and create an artificial feedback effect on $Y_{e}$.\\

\begin{figure}[h]
 \begin{center}
\rotatebox{-90}{
\includegraphics[angle=0,bb=-5 60 580 5500,clip,scale=0.35]{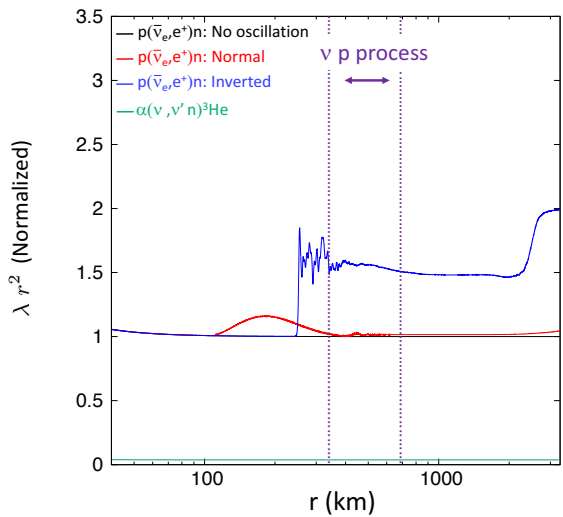}
}
\caption{The evolution of normalized $\lambda r^{2}$ in the early outflow model ($t=0.6$ s) where $\lambda$ represents the reaction rate of the charged current reaction $\bar{\nu}_{e}+p\to e^{+}+n$ or the neutral current reaction $\alpha+\nu\to\ ^{3}\mathrm{He}+n+\nu^{\prime}$. The $\nu p$ process is active in the region $r\sim350-680$ km.
} 
\label{FIG:make_reaction_rate_0.6} 
 \end{center}
 \end{figure}

 \ The flavor transitions of energetic $\bar{\nu}_{e}$s at the radius of $r\sim350-680$ km play a crucial role in the enhancement of the $\nu p$ process which results in the production of more abundant $p$-nuclei. The $\nu p$ process happens through $(n,p)$ and $(p,\gamma)$ reactions in proton-rich wind trajectories  ($Y_{e}>0.5$). The $\nu p$ process occurs during $T\sim2-3\times10^{9}$ K \cite{Wanajo 2006} which corresponds to $r\sim350-680$ km in our proton-rich outflows. 
Most of free neutrons are produced by $\bar{\nu}_{e}+p\to e^{+}+n$. In addition, the reaction $\alpha+\nu\to\ ^{3}\mathrm{He}+n+\nu^{\prime}$ also supplies abundant free neutrons after $\alpha$-particles become dominant species ($T<6\times10^{9}$ K).\\

\ The modification of neutrino energy spectra due to collective oscillations affects the neutrino induced reaction rates. There is no oscillation effect in $\alpha$($\nu$,$\nu^{\prime}$$n$)$^{3}\mathrm{He}$ because this is a neutral current reaction. On the other hand, the reaction rate of $p$($\bar{\nu}_{e}$,$e^{+}$)$n$ can probe the oscillation effects as shown in Fig.\ref{FIG:make_rho_be_evolution_3_m_0.6}(a)(b) because this quantity is derived by the integration of $\bar{\rho}_{ee}(r,E,\theta_{p})$:
\begin{eqnarray}
\label{eq:reaction rate of nu_be absorption}
&& \lambda_{\bar{\nu}_{e}} = \int\mathrm{d}E\ \mathrm{d}\cos\theta_{p} \nonumber \\
&&\sum_{\alpha=e,\mu,\tau}\frac{L_{\bar{\nu}_{\alpha}}}{2\pi R_{\nu}^{2}\average{E_{\bar{\nu}_{\alpha}}}}f_{\bar{\nu}_{\alpha}}(E)\bar{\rho}_{ee}(r,E,\theta_{p})\sigma_{\bar{\nu}_{e}}(E) .
\end{eqnarray}
\ Fig.\ref{FIG:make_reaction_rate_0.6} displays the evolution of normalized $\lambda r^{2}$ where $\lambda$ is the reaction rate of $p$($\bar{\nu}_{e}$,$e^{+}$)$n$ or $\alpha$($\nu$,$\nu^{\prime}$$n$)$^{3}\mathrm{He}$. 
Without neutrino oscillations, the reaction rate decreases as $\lambda\propto 1-\sqrt{1-(R_{\nu}/r)^{2}}\sim 1/2(R_{\nu}/r)^{2}\ (r>>R_{\nu})$. The value of $\lambda r^{2}$ is normalized by the final $\lambda_{\bar{\nu}_{e}}r^{2}$ calculated in no oscillation case (black curve in Fig.\ref{FIG:make_reaction_rate_0.6}).\\

\  In normal mass hierarchy (the red curve in Fig.\ref{FIG:make_reaction_rate_0.6}), collective neutrino oscillations enhance the value of $\lambda_{\bar{\nu}_{e}}$ around $r=150$ km where almost all nuclides are in quasi-statistical equilibrium (QSE) \cite{Meyer 1998}. In the QSE state ($T\sim3\mathrm{-}5\times10^{9}$ K), all nuclear abundance ratios are determined by the temperature, density, $Y_{e}$ and a small amount of heavy nuclei $Y_{h}$ in the system. The feedback effect of neutrino oscillations on $Y_{e}$ is negligible (see the discussion in section \ref{sec:Setup for numerical simulations}). Therefore, the increased reaction rate does not affect the nucleosynthesis strongly in this region. Seeds nuclei for heavy elements such as $^{56}\mathrm{Ni}$, $^{60}\mathrm{Zn}$ and $^{64}\mathrm{Ge}$ are synthesized by $\alpha$-capture reactions before the $\nu p$ process is ignited.\\

\ In inverted mass hierarchy (the blue curve in Fig.\ref{FIG:make_reaction_rate_0.6}), the value of $\lambda_{\bar{\nu}_{e}}$ is increased by collective neutrino oscillations and its high value is maintained in $r\sim350-680$ km, which shows the oscillation effects enhance the $\nu p$ process successfully. The enhancement of $\lambda_{\bar{\nu}_{e}}$ is mainly due to the flavor transitions in energetic antineutrinos because of the energy dependence of the cross section $\sigma_{\bar{\nu}_{e}}(E)\propto(E/\mathrm{MeV}-1.293)^{2}$. Therefore, the contribution from increasing high energy electron antineutrinos ($E>E_{c2}^{(e)}$) is larger than that of decreasing intermediate electron antineutrinos ($E_{c1}^{(e)}<E<E_{c2}^{(e)}$) in Fig.\ref{FIG:make_spec_0.6}(e). After the $\nu p$ process has terminated, $\beta^{+}$ decays and $(n,\gamma)$ dominate the nuclear reactions inside the neutrino-driven wind. The MSW effects increase the value of $\lambda_{\bar{\nu}_{e}}$ around $r=2000$ km. These oscillation effects are negligible in the nucleosynthesis because neutrino-induced reactions fail to produce much free neutrons for the subsequent $(n,\gamma)$ and $(n,p)$ reactions in this outer region.


 \begin{figure*}
 \begin{center}
 \rotatebox{-90}{
 \includegraphics[angle=0,bb=-5 0 530 840,clip,scale=0.5]{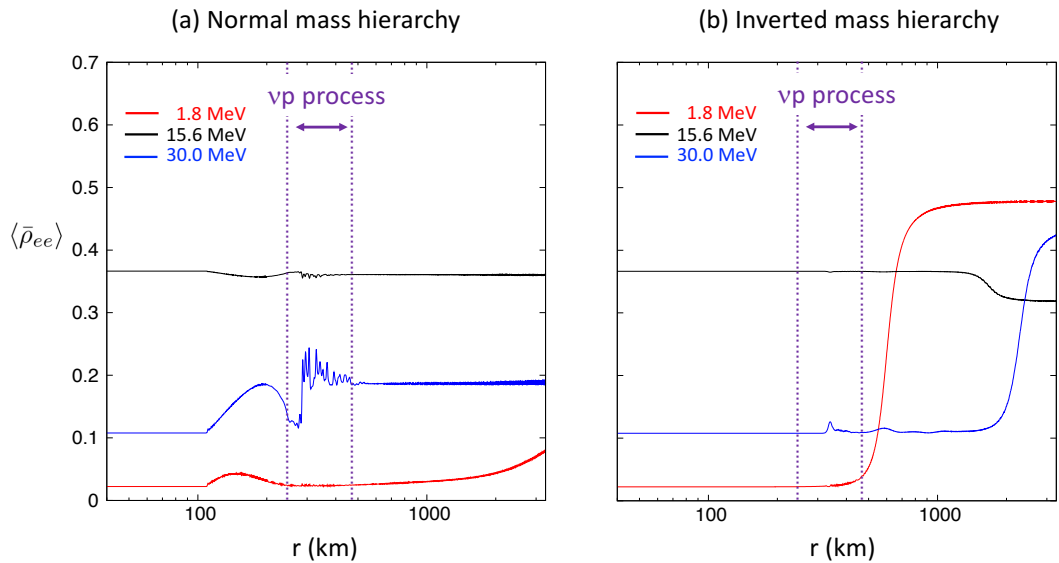}
 }
 \caption{The evolution of \average{\bar{\rho}_{ee}(r,E)} in normal (a) and inverted (b) mass hierarchies using the later wind trajectory ($t=1.1$ s). The $\nu p$ process occurs during $r\sim245-470$ km.}\label{FIG:make_rho_be_evolution_3_m}
 \end{center}
 \end{figure*}

  \subsection{Later neutrino-driven wind $(t=1.1$ s$)$} 
\label{subsec:The later trajectory $(t=1.1$ s$)$}
Collective neutrino oscillations are very sensitive to the ratio of neutrino number fluxes between all species of neutrinos \cite{Mirizzi 2011}. In the cooling phase, the neutrino luminosity is decreasing, which changes the neutrino number fluxes resulting in the variety of collective neutrino oscillations. In normal mass hierarchy, sharp flavor transitions of energetic antineutrinos around $r=280$ km have significant effects on the $\nu p$ process.\\

\begin{figure}
 \begin{center}
 \rotatebox{0}{
\includegraphics[angle=0,bb=30 0 1930 800,clip,scale=0.45]{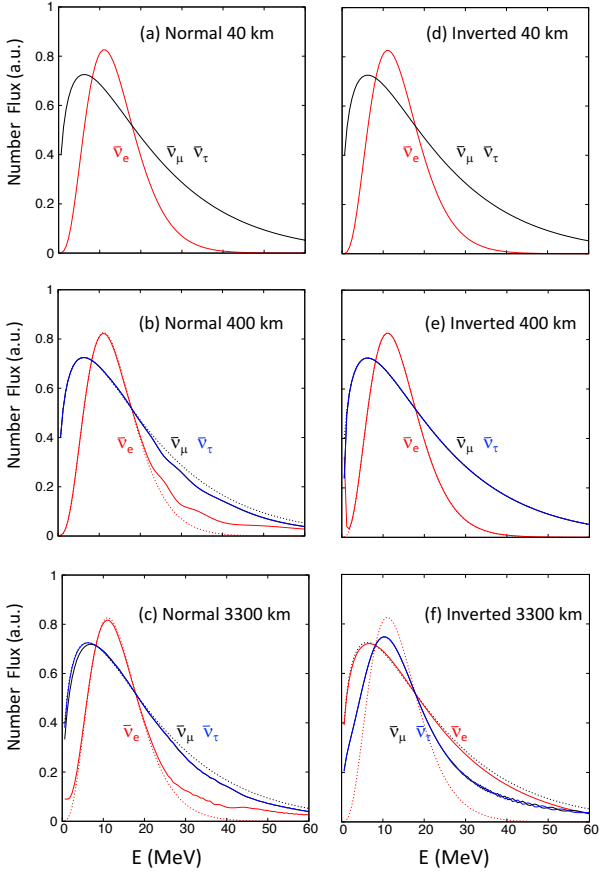}
}
 \caption{The evolution of energy spectra of antineutrinos for the later wind trajectory ($t=1.1$ s) as in Fig.\ref{FIG:make_spec_0.6}. The spectral crossing points are $E_{c1}^{(l)}=8.2$ MeV and $E_{c2}^{(l)}=17.8$ MeV.}\label{FIG:make_spec}
 \end{center}
 \end{figure}

In normal mass hierarchy (Fig.\ref{FIG:make_rho_be_evolution_3_m}(a)), flavor transitions occur in $r\sim110$ km. Then, antineutrinos gradually come back to their original flavors. This oscillation behavior is similar to that of early trajectory (Fig.\ref{FIG:make_rho_be_evolution_3_m_0.6}(a)). However, in the later trajectory, a sharp flavor transition occurs raising the value of $\average{\bar{\rho}_{ee}(r,E)}$ around 280 km for high energy antineutrinos as shown, for example, by the 30 MeV antineutrinos in Fig.\ref{FIG:make_rho_be_evolution_3_m}(a). The spectra of antineutrinos in $r=400$ km are shown in Fig.\ref{FIG:make_spec}(b) which reflects these sharp flavor transitions in high energy antineutrinos whose energy is larger than the second spectral crossing point $E_{c2}^{(l)}=17.8$ MeV. After that, low energy antineutrinos are transformed to the vacuum mass eigenstates adiabatically because of the matter effects as discussed in the early trajectory.\\

\ In inverted mass hierarchy (Fig.\ref{FIG:make_rho_be_evolution_3_m}(b)), collective neutrino oscillations start around 330 km, but oscillation amplitudes of antineutrinos are highly suppressed by the multi-angle decoherence. The effects of collective neutrino oscillations on the antineutrino spectra are negligible as shown in Fig.\ref{FIG:make_spec}(e). After the collective neutrino oscillations cease, antineutrinos undergo MSW resonances in outer regions. As shown in Fig.\ref{FIG:make_rho_be_evolution_3_m}(b), the resonance point depends on the energy of the antineutrino because the value of the critical electron density is proportional to $E^{-1}$. The spectral split of low energy antineutrinos ($E\sim1$ MeV) in Fig.\ref{FIG:make_spec}(e) is caused by the MSW resonance. In the later explosion phase, the electron density inside the outflow decreases more rapidly compared with that in the early phase. Therefore, the MSW resonance occurs in the later outflow while the electron density of the early wind trajectory can not decrease down to the critical values. In outer region, high energy $\bar{\nu}_{e}$ also begins to transform to $\bar{\nu}_{y}$ resulting in dramatic spectral swaps in antineutrino spectra as shown in  Fig.\ref{FIG:make_spec}(f).\\



 

  \begin{figure}[h]
 \begin{center}
\rotatebox{-90}{
\includegraphics[angle=0,bb=-5 60 580 5500,clip,scale=0.35]{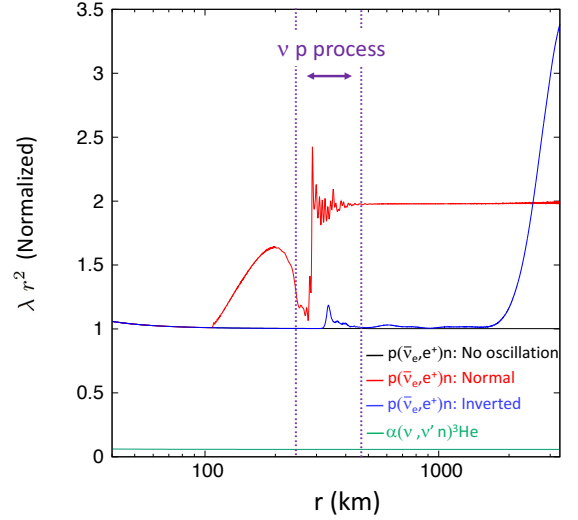}
}
\caption{The evolution of normalized $\lambda r^{2}$ in the later outflow model ($t=1.1$ s) as in Fig.\ref{FIG:make_reaction_rate_0.6}. The $\nu p$ process occurs in the interval $r\sim245-470$ km.
} 
\label{FIG:make_reaction_rate} 
 \end{center}
 \end{figure}

\ Fig.\ref{FIG:make_reaction_rate} represents the evolution of neutrino-induced reaction rates in the later trajectory ($t=1.1$ s). In this outflow model, the gas temperature immediately decreases down to $T\sim2-3\times10^{9}$ K ($r\sim245-470$ km) where large amount of heavy $p$-nuclei and their seed nuclides are synthesized through the $\nu p$ process.\\

\ In normal mass hierarchy (red curve in Fig.\ref{FIG:make_reaction_rate}), the early enhancement of $\lambda_{\bar{\nu}_{e}}$ near the onset of collective neutrino oscillations ($r\sim110$ km) can not contribute to the nucleosynthesis as discussed in the early wind model. However, the enhancement of $\lambda_{\bar{\nu}_{e}}$ around $280$ km makes a remarkable influence on the nucleosynthesis. The sharp flavor transitions around $280$ km in energetic antineutrinos raise the value of $\lambda_{\bar{\nu}_{e}}$ by a factor two. The raised value of $\lambda_{\bar{\nu}_{e}}$ is kept up until the $\nu p$ process freezes out inside the outflow. Therefore, the $\nu p$ process is enhanced successfully resulting in the productions of more abundant $p$-nuclei.\\

\ On the other hand, in inverted mass hierarchy (blue curve in Fig.\ref{FIG:make_reaction_rate}), the effect of neutrino oscillations on nucleosynthesis is not significant because dramatic flavor transitions do not occur in 
the region 
$r\sim245-470$ km. Even though $\lambda_{\bar{\nu}_{e}}$ increases later because of the MSW resonances and finally exceeds the corresponding value in the normal mass hierarchy, the $\nu p$ process has already finished, and few free neutrons are produced in rapidly expanding outflows at high wind velocity $v(r)\sim3\times10^{9}$cm/s. Therefore, oscillation effects are not expected to significantly affect the $\nu p$ process and neutron-capture reactions.\\ 

\subsection{The abundances of $p$-nuclei produced in neutrino-driven winds}
 \label{subsec:The abundances of $p$-nuclei produced in neutrino-driven winds}

 \begin{figure*}
\begin{center}
\rotatebox{-90}{
 \includegraphics[angle=0,bb=50 0 450 3500,clip,scale=0.63]{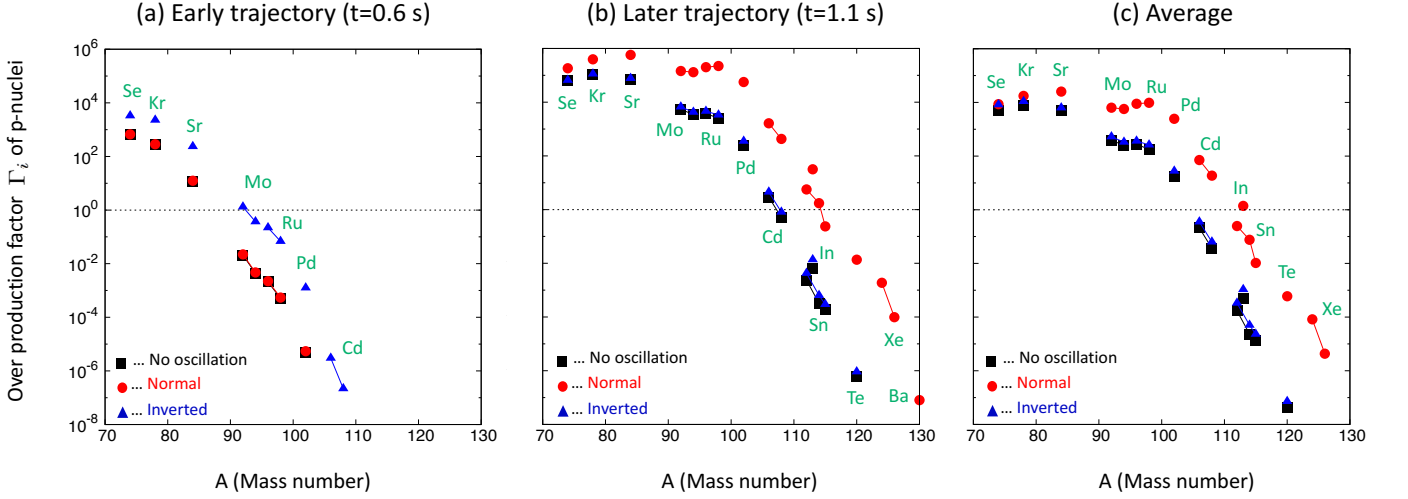}
 }
\caption{The overproduction factors of $p$-nuclei $\Gamma_{i}$ in the early trajectory (a), later trajectory (b). The averaged value of overproduction factor (c) is obtained by using Eq.(\ref{eq:overproduction_average}) and (\ref{eq:f}).}\label{FIG:make_ratio_sum}
 \end{center}
\end{figure*}
  

In this section, we discuss influence of collective neutrino oscillations on abundances of $p$-nuclei produced through the $\nu p$ process nucleosynthesis in both early ($t=0.6$ s) and later ($t=1.1$ s) wind trajectories. In the early outflow, oscillation effects are prominent in inverted mass hierarchy. On the other hand, in the later trajectory, heavy $p$-nuclei are highly enhanced in normal mass hierarchy. Finally, abundances are to be averaged over the different wind contributions.\\



\ Fig.\ref{FIG:make_ratio_sum} represents the overproduction factors of $p$-nuclei. The overproduction factor for the nucleus $i$ is defined by
\begin{equation}
\label{eq:overproduction}
\Gamma_{i}=\frac{X_{i}}{X_{i, \mathrm{solar}}}/\frac{X_{^{56}\mathrm{Fe}}}{X_{^{56}\mathrm{Fe}, \mathrm{solar}}},
\end{equation}
where $X_{i}$ and $X_{i,\mathrm{solar}}$ are the mass fractions of nucleus $i$ in the wind trajectory and in the solar system \cite{Lodders 2003}, respectively. $X_{i}$ is derived by carrying out the nucleosynthesis calculation until all nuclear reactions freeze out. In case of $\Gamma_{i}>1$, large amounts of nucleus $i$ are produced which are enough to explain the solar abundance of nucleus $i$ if we assume that $^{56}\mathrm{Fe}$ in the solar system is produced only by this wind trajectory. Very large $\Gamma_{i}$ does not make trouble in such an interpretation as to be discussed below in Eq.(\ref{eq:overproduction_average}) and in the next section.\\

\ The over production factors of $p$-nuclei in the early trajectory model are shown in Fig.\ref{FIG:make_ratio_sum} (a). These results reflect the behavior of collective neutrino oscillations and their effects of the $\lambda_{\bar{\nu}_{e}}$ discussed in section \ref{subsec:The early trajectory $(t=0.6$ s$)$}. In normal mass hierarchy, oscillation effects hardly contribute to the production of $p$-nuclei as implied in Fig.\ref{FIG:make_reaction_rate_0.6}, so that the value of $\Gamma_{i}$ is similar to that of no oscillation case. In inverted mass hierarchy, however, $p$-nuclei are increased by up to $\sim10^{2}$ times owing to the enhancement of $\lambda_{\bar{\nu}_{e}}$ during $r\sim350-680$ km. Heavy $p$-nuclei tends to be created more abundantly because high $\lambda_{\bar{\nu}_{e}}$ supplies more free neutrons for subsequent $(n,p)$ reactions on heavy elements as discussed in Ref. \cite{Martinez 2011}. The $p$-nuclei are synthesized up to $^{106}\mathrm{Cd}$ and $^{108}\mathrm{Cd}$ even when oscillation effects are taken into account. In the early trajectory, temperature decreases slowly compared with the time scale of $\alpha$-capture reactions \cite{Otsuki 2000}, so that more $^{56}\mathrm{Ni}$ are synthesized before the $\nu p$ process takes place resulting in the small production of heavier elements.\\



\ Fig.\ref{FIG:make_ratio_sum}(b) represents the production of $p$-nuclei in the later outflow model. In normal mass hierarchy, more $p$-nuclei are synthesized by the collective neutrino oscillations. These oscillation effects allow the nuclear flow to reach heavier $p$-nuclei like $^{124}\mathrm{Xe}$, $^{126}\mathrm{Xe}$ and $^{130}\mathrm{Ba}$ on the chart of nuclides which fail to be synthesized in no oscillation case. Overproduction factors of these $p$-nuclei are extremely enhanced by up to $\sim10^{4}$ times. The amount of enhancement in our model is quite larger than that of in Ref. \cite{Martinez 2011} (up to $\sim 20$ times). Our initial neutrino parameters are such that there is a large excess of $\nu_{\beta}$ over $\bar{\nu}_{e}$ in $E>E_{c2}^{l}$, which creates favorable conditions for the enhancement of energetic $\bar{\nu}_{e}$ through collective neutrino oscillations. The increased flux of $\bar{\nu}_{e}$ at high energy region results in the large overproduction factor in our model due to the energy dependence of the cross section $\sigma_{\bar{\nu}_{e}}(E)\propto(E/\mathrm{MeV}-1.293)^{2}$. In inverted mass hierarchy, oscillation effects on the $\nu p$ process are small and fail to increase $p$-nuclei sufficiently even though MSW resonances cause significant enhancement of $\lambda_{\bar{\nu}_{e}}$ after the $\nu p$ process, as already discussed in section \ref{subsec:The later trajectory $(t=1.1$ s$)$}. The nucleosynthesis in no oscillation case also fails to synthesize heavier $p$-nuclei although lighter $p$-nuclei such as $^{74}\mathrm{Se}, ^{78}\mathrm{Kr}$ and $^{84}\mathrm{Sr}$ are produced.\\

\ We average overproduction factors of $p$-nuclei using both early wind models, as represented in our fiducial model of $t=0.6$ s, and later ones, as represented in our fiducial model of $t=1.1$ s. We can roughly regard this quantity as the overproduction factor of the total ejecta in cooling phase. The averaged overproduction factor $\average{\Gamma_{i}}$ is defined by
\begin{equation}
\label{eq:overproduction_average}
\average{\Gamma_{i}}=(1-f)\ \Gamma_{i}|_{\mathrm{early}}+f\ \Gamma_{i}|_{\mathrm{later}},
\end{equation}
where $\Gamma_{i}|_{\mathrm{early}}$ and $\Gamma_{i}|_{\mathrm{later}}$ are overproduction factors of nucleus $i$ in the early and later winds respectively. The ratio $f$ is the mass weight for the average determined by
\begin{equation}
\label{eq:f}
f=\frac{\Delta M_{^{56}\mathrm{Fe}}|_{\mathrm{later}}}{\Delta M_{^{56}\mathrm{Fe}}|_{\mathrm{early}}+\Delta M_{^{56}\mathrm{Fe}}|_{\mathrm{later}}},
\end{equation}
where $\Delta M_{^{56}\mathrm{Fe}}|_{\mathrm{early}}$ and $\Delta M_{^{56}\mathrm{Fe}}|_{\mathrm{later}}$ are the ejected mass of $^{56}\mathrm{Fe}$ in the early phase ($0.6$ s $<t<1.1$ s) and the later phase ($t>1.1$ s). Table \ref{tab:average overproduction} shows the ratio $f$ and ejected iron mass. $\Delta M_{^{56}\mathrm{Fe}}|_{\mathrm{later}}$ is estimated assuming that the contribution of the later phase is effective up to $t\sim3$ s because of the small mass ejection after $t>3$ s \cite{Pllumbi 2015}.\\

\ The averaged overproduction factor $\average{\Gamma_{i}}$ is shown in Fig.\ref{FIG:make_ratio_sum} (c). In lighter $p$-nuclei such as $^{74}\mathrm{Se}, ^{78}\mathrm{Kr}$ and $^{84}\mathrm{Sr}$, the hierarchy difference is reduced because of the contributions from the early phase in inverted hierarchy case. In heavier $p$-nuclei ($A>92$), the contribution from the later phase is dominant despite the small value of $f$. Heavy elements are more efficiently synthesized in the later phase because of the small dynamical time scale of gas temperature \cite{Otsuki 2000}. In addition, the high electron fraction $Y_{e}\sim0.59$ which causes abundant target protons for $\bar{\nu}_{e}+p\to e^{+}+n$ promotes the $\nu p$ process actively, resulting in the high values of $\Gamma_{i}$ for the $p$-nuclei.\\


\begin{table*}[htb]
\begin{center}
\small
\begin{tabular}{|cccc|} \hline
&$\Delta M_{^{56}\mathrm{Fe}}|_{\mathrm{early}}$&$\Delta M_{^{56}\mathrm{Fe}}|_{\mathrm{later}}$&\ \ f\\ 
&($\times10^{-6} M_{\odot}$)&($\times10^{-6} M_{\odot}$)&\ \ ($\times10^{-2}$) \\ \hline
No ocillations&68.0&5.29&\ \ 7.2\\
Normal&68.0&3.11&\ \ 4.4\\ 
Inverted&60.8&5.17&\ \ 7.8\\ \hline
\end{tabular}
\end{center}
\small
\caption{The amount of ejected $^{56}\mathrm{Fe}$ and the ratio $f$ in each hierarchies.}\label{tab:average overproduction}
\end{table*}

 \section{Summary and Discussions}
 \label{sec:Summary and Discussion}

\ We studied three flavor multi-angle collective neutrino oscillations together with nucleosynthesis network calculations using two proton-rich neutrino-driven winds at $t=0.6$ s and $1.1$ s after core bounce obtained by the 1D explosion simulation model. We choose these outflows as representatives of early and later trajectories in cooling phase. \\

\ In the early wind trajectory ($t=0.6$ s), the number flux of energetic electron antineutrinos is increased by the collective neutrino oscillations in inverted mass hierarchy during $r\sim350-680$ km where the $\nu p$ process nucleosynthesis takes place. High energy electron antineutrinos play a significant role in the $\nu p$ process because of the large cross section in Eq.(\ref{eq:cross section nube}). These oscillation effects promote the $\nu p$ process actively producing more abundant $p$-nuclei by up to $10^{2}$ times larger than those in no oscillation case. \\



  
 \ On the other hand, in the later trajectory ($t=1.1$ s), we find that the $\nu p$ process is dramatically enhanced in normal mass hierarchy by sharp flavor transitions in $r\sim280$ km which increase energetic electron antineutrinos. In the literature, it is reported many times that neutrino self interactions cause neutrino spectral swaps in inverted mass hierarchy \cite{Fogli 2007,Duan 2010}. However, such effects are also reported for the normal mass hierarchy \cite{Dasgupta 2010} in a set of initial condition where number fluxes of non-electron (anti)neutrino flavors are larger than that of electron (anti)neutrino, which is also the case in our simulation. The enhanced $\nu p$ process allows the value of overproduction factor of $p$-nuclei $\Gamma_{i}$ to be raised by $\sim10-10^{4}$ times. The results highly depend on the initial neutrino parameters on the surface of neutrino sphere which are shown in Table \ref{tab:multi-angle parameters}. The dramatic enhancement of $p$-nuclei is partially due to the large excess of non-electron antineutrinos over electron antineutrinos in high energy region.\\


\ Our results indicate the necessity of incorporating the effects of collective neutrino oscillations for precise $\nu p$ process nucleosynthesis calculations in wind trajectories. The fact that the overproduction factors of heavy $p$-nuclei are dominated by the later wind reduces the model dependence of our results because our treatment which assumes steady state outflows is applicable very well to later wind trajectories. Furthermore, our finding also suggests that such precise theoretical studies of $\nu p$ process nucleosynthesis can potentially identify the still unknown origin of the solar $^{92,94}\mathrm{Mo}$ and $^{96,98}\mathrm{Ru}$ \cite{Woosley 1978,Hayakawa 2008}.\\

 \ We calculate the averaged overproduction factor of $p$-nuclei $\average{\Gamma_{i}}$ by using only two wind trajectories at $t=0.6$ s and $1.1$ s. More quantitative discussion about the nucleosynthesis is desirable by using many more wind trajectories beyond $t=1.1$ s which were ignored in the present calculation due to limited computational resources. In addition, the contributions of the outer Si-burning layer are necessary to obtain the total abundance of these nuclides produced in this explosion model. The net overproduction factors would be $\Gamma_{i}\sim1$ if the solar abundances of $p$-nuclei are explained successfully in the supernova model. In the present calculation,  $\average{\Gamma_{i}}$ for $^{92,94}\mathrm{Mo}$ and $^{96,98}\mathrm{Ru}$ take large values $\sim10^{4}$ in normal mass hierarchy. Taking into account the contributions of all other ejecta, the values of $\average{\Gamma_{i}}$ will be lowered by several orders because the large amount of  $^{56}\mathrm{Fe}$ is produced there. In our rough estimate assuming the amount of total $^{56}\mathrm{Ni}$ ejecta $M_{^{56}\mathrm{Ni}}=0.07 M_{\odot}$, the values of $\average{\Gamma_{i}}$ in the present study decrease by three orders of magnitude. \\
 
 

\ The caveat of this
study is the uncertainty of  neutrino parameters describing neutrino spectra.  Both collective neutrino
oscillations and explosive nucleosynthesis highly depend on the initial neutrino
parameters.
  If the differences between luminosities and energies of different neutrino species are very small, oscillation effects on $\lambda_{\bar{\nu}_{e}}$ are negligible. In our explosion model, the value of $\average{E_{\nu_{\beta}}}$ may decrease and approach to that of $\average{E_{\bar{\nu}_{e}}}$ if we included neutral current reactions discussed in Ref. \cite{Muller 2012}. Such modifications may lower the initial number flux of $\nu_{\beta}$ in high energy region reducing the enhancement of $\lambda_{\bar{\nu}_{e}}$ as shown in Ref. \cite{Wu 2015}. However, note that nucleon-nucleon correlation may increase neutrino mean energies \cite{Horowitz 2017,Burrows 2017}.\\
  
  \ The $\nu p$ process depends not only on initial neutrino parameters but also on hydrodynamic quantities. In particular, the wind velocity $v(r)$ is important for the $\nu p$ process nucleosynthesis as discussed in our preliminary study \cite{Sasaki 2017}. Free neutrons supplied by $p$($\bar{\nu}_{e}$,$e^{+}$)$n$ from $r$ to $r+\Delta r$ are represented by
$\Delta Y_{n}|_{\mathrm{cc}}=\lambda_{\bar{\nu}_{e}}Y_{p}\Delta r/v(r)$
 where $Y_{p}$ is the abundance of free protons. $\Delta Y_{n}|_{\mathrm{cc}}$ can be amplified easily in a slower wind trajectory leading to large variation of the effects of collective neutrino oscillations. 
Therefore, a comprehensive and systematic study of hydrodynamic quantities as well as initial neutrino parameters is desirable in order to better understand the behavior of collective neutrino oscillations and the properties of nucleosynthesis in neutrino-driven winds.\\






\section*{Acknowledgement}
We thank M. Kusakabe for useful discussion on
nucleosynthesis and for providing neutrino-induced reaction cross
sections. This work was supported in part by Grants-in-Aid for
Scientific Research of JSPS  (26105517, 24340060, 26870823, 15H01039, 15H00789, 15H03665, 17K05457, 17H01130, 17K14306 and 17H05206), in part by the US National Science
Foundation Grant No. PHY-1514695, and in part by Turkish Scientific and
Technological Research Council Project No. 115F548.


\end{document}